\documentclass[sigconf,natbib=false,authorversion]{acmart}

\usepackage{graphicx}
\usepackage{subcaption}
\expandafter\def\csname ver@subfig.sty\endcsname{}

\usepackage{xcolor}
\usepackage{url}
\usepackage{graphicx}
\usepackage{xspace}
\usepackage{subfig}
\usepackage{setspace}

\makeatletter
\renewcommand\paragraph{\@startsection{paragraph}{4}{\parindent}%
  {0pt}
  {-\parindent}
  {\ACM@NRadjust{\normalfont\bfseries\@adddotafter}}}
\makeatother

\newcommand{\ie}{i.e.,\xspace}
\newcommand{\eg}{e.g.,\xspace}

\newcommand{\cf}{cf.\xspace}
\newcommand{\etal}{et al.\xspace}

\usepackage[inline]{enumitem}
\setlist[itemize]{label=\textbullet,topsep=0pt,leftmargin=1em}
\setlist[enumerate]{label=(\alph*)}

\newcommand{\stepnumber}[2][]{\raisebox{.5pt}{\textcircled{\raisebox{-.5pt}{\footnotesize#2\raisebox{1.5pt}{\tiny #1}}}}}
\newcommand{\stepnumberspace}[3][]{\stepnumber[#1]{#2}{\kern 0.25em}#3}

\newcommand{\name}{ACRP\xspace}

\copyrightyear{2023}
\acmYear{2023}
\setcopyright{rightsretained}
\acmConference[CCS '23] {Proceedings of the 2023 ACM SIGSAC Conference on Computer and Communications Security}{November 26--30, 2023}{Copenhagen, Denmark.}
\acmBooktitle{Proceedings of the 2023 ACM SIGSAC Conference on Computer and Communications Security (CCS '23), November 26--30, 2023, Copenhagen, Denmark}
\acmPrice{15.00}
\acmPrice{}
\acmISBN{979-8-4007-0050-7/23/11}
\acmDOI{10.1145/3576915.3624367}

\RequirePackage[
    datamodel=acmdatamodel,
    style=acmnumeric,
    maxbibnames=2,
    giveninits=true,
    doi=false,
    isbn=false,
]{biblatex}

\begin{document}

\title{Poster: Accountable Processing of Reported Street Problems}

\author{Roman Matzutt}
\email{matzutt@comsys.rwth-aachen.de}
\orcid{0000-0002-4263-5317}
\affiliation{%
    \institution{RWTH Aachen University}
    \streetaddress{Ahornstraße 55}
    \city{Aachen}
    \country{Germany}
    \postcode{52074}
}

\author{Jan Pennekamp}
\email{jan.pk@comsys.rwth-aachen.de}
\orcid{0000-0003-0398-6904}
\affiliation{%
    \institution{RWTH Aachen University}
    \streetaddress{Ahornstraße 55}
    \city{Aachen}
    \country{Germany}
    \postcode{52074}
}

\author{Klaus Wehrle}
\email{wehrle@comsys.rwth-aachen.de}
\orcid{0000-0001-7252-4186}
\affiliation{%
    \institution{RWTH Aachen University}
    \streetaddress{Ahornstraße 55}
    \city{Aachen}
    \country{Germany}
    \postcode{52074}
}

\renewcommand{\shortauthors}{Roman Matzutt, Jan Pennekamp, \& Klaus Wehrle}

\begin{abstract}
    Municipalities increasingly depend on citizens to file digital reports about issues such as potholes or illegal trash dumps to improve their response time.
    However, the responsible authorities may be incentivized to ignore certain reports, \eg when addressing them inflicts high costs.
    In this work, we explore the applicability of blockchain technology to hold authorities accountable regarding filed reports.
    Our initial assessment indicates that our approach can be extended to benefit citizens and authorities in the future.
\end{abstract}



\maketitle

\section{Introduction}%
\label{sec:introduction}

The ongoing digitization of everyday life has benefited the emergence of crowdsensing applications, which allow monitoring a large area using customer devices~\cite{2015_guo_mobile_crowdsensing}.
Examples of crowdsensing include
    environmental monitoring~\cite{2009_mun_peir},
    traffic anomaly detection~\cite{2013_pan_crowdsensing_traffic},
    and contact tracing during the coronavirus pandemic~\cite{2021_munzert_contact_tracing}.

Another application for crowdsensing is the distributed monitoring of \emph{street problems}, such as potholes or other traffic obstructions, illegal trash dumps, graffiti, or other damages.
Relying on citizens to report the issues they observe promises to
\begin{enumerate*}
    \item increase the covered area,
    \item reduce the associated costs for city officials, and
    \item enable a citizen-centered prioritization of identified issues.
\end{enumerate*}

However, current corresponding platforms illustrate that this application lacks a clear candidate for operating the required infrastructure in a binding and transparent manner.
On the one hand, NGOs may operate corresponding platforms, potentially with a special focus such as bike lane safety~\cite{2007_cyclinguk_fillthathole} or tracking anonymous reports of sexual violence to identify and improve unsafe areas~\cite{2012_reddotfoundation_safecity}.
Unfortunately, constituting separate organizations implies that reports are not binding for the affected administrations that would need to handle those reports.
On the other hand, governments or municipalities may operate corresponding platforms themselves, \eg using FixMyStreet Platform~\cite{2012_fixmystreet_platform}.
However, even if the official government bodies claim to commit to handling reports to those platforms, giving them complete control over the reports can undermine any perceived accountability via technical means.
For instance, if an issue is deemed too expensive to fix, the municipality has a financial incentive to hide the issue from the broader public.

In this work, we propose an initial framework that resolves this tension and enables citizens to file their reports while ensuring that the responsible authorities can be held accountable for neglecting them.
Our framework, the \emph{accountable city report platform (\name)}, traces reports on a consortium blockchain that is jointly operated by independent municipalities and, optionally, additional NGOs.
This way, reports are recorded immutably, \ie municipalities can neither hide nor modify any report later on.
The potential use of blockchain technology for active citizen participation has been explored before~\cite{2020_bendhaou_smart_city}, but, to the best of our knowledge, these use cases do no currently consider data submitted by the citizens themselves.

\section{Overview and Value Proposal}%
\label{sec:problem-statement}

\begin{figure}[b]
    \centering
    \includegraphics{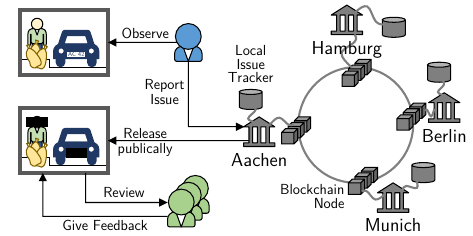}
    \caption{
        \name ensures that citizen-reported street problems are handled in an accountable manner.
    }%
    \label{fig:overview}
\end{figure}

Figure~\ref{fig:overview} illustrates our target architecture:
Citizens are enabled to file digital reports of street problems that require attention from the responsible municipality.
This report contains all information required to address the issue.
The municipality keeps track of reported issues locally, but tracks its state on a consortium blockchain.
The blockchain is operated by a group of independent municipalities and, optionally, NGOs, but must be publicly readable and the operating nodes must accept citizen-submitted data.
Finally, the report is released to the public to establish public accountability.

In the following, we discuss that this approach can increase both accountability and data quality but also raises new challenges.

\subsection{Benefits}%
\label{sec:problem-statement:benefits}

The blockchain-backed and publicly observable approach of \name promises several advantages for both citizens and municipalities.

\paragraph{Accountability}

The underlying blockchain can assert citizens that a trace of their filed reports is immutably recorded, \ie the responsible authorities cannot deny having received any recorded report.
This traceability extends to any subsequent handling of the report.
As a result, reports cannot be altered or removed without due reason, which establishes the required accountability.

\paragraph{Reduced Duplicates}

Blockchain-recorded data traces enable a globally agreed-upon view on the existence and current state of reports.
Together with additional similarity checks, citizens and authorities can identify and merge potential duplicates more easily, reducing the workload for all involved entities.

\paragraph{Weighted Prioritization}

Adding features such as up-voting or short comments in addition to the now-achievable transparency and deduplication further helps authorities assess and compare the \emph{relevance} of identified street issues.
Namely, citizens can better express the importance of individual issues, which allows authorities to optimize the achievable benefit for a given limited budget.

\paragraph{Citizen Mobility}

Finally, a shared infrastructure facilitates mobile citizens, such as commuters or tourists.
These groups would benefit from a unified interface to file reports about street problems relevant to their mobility patterns across different municipalities.

Overall, our approach promises benefits for citizens and municipalities.
However, recording reports on a publicly readable blockchain comes with severe challenges, as we outline in the following.

\vspace{-0.7em}
\subsection{Challenges}%
\label{sec:problem-statement:challenges}

Despite the outlined benefits, our approach also creates new challenges \name must ultimately overcome for maximum acceptance.

\paragraph{Reporter and Citizen Privacy}

Blockchain-recorded data has the potential to compromise the privacy of citizens, both directly~\cite{2018_matzutt_contents} and indirectly~\cite{2013_reid_anonymity_analysis}.
Most notably, reports should include a photo to help authorities assess the legitimacy and extent of the issue.
However, these photos may also include, \eg bystanders' faces or license plates of nearby parked cars.
For instance, Google's Street View raised various privacy concerns~\cite{2011_rakower_gsv_privacy} and Google had to start blurring such information on a large scale~\cite{2009_frome_gsv_blur}.
Unfortunately, directly transferring this modification approach to blockchain-backed reports would interfere with the desired report accountability.
Finally, reporters may indirectly disclose information about their individual points of interest via their reports' locations~\cite{2017_ziegeldorf_tracemixer}.

\paragraph{Data Quality}

Reports are only valuable if they describe the issue at hand accurately enough so that the responsible authorities can take action.
In addition to the privacy concerns discussed above, the \emph{credibility} of submitted reports has to be ensured~\cite{2019_luo_data_quality}.
Since \name must provide an open platform for citizens, it must be able to handle reports that are
\begin{enumerate*}
    \item too low-quality to be actionable,
    \item forged to provoke an unnecessary reaction with associated costs, and
    \item ``garbage'' reports, \eg containing spam or illicit content.
\end{enumerate*}

\paragraph{Usability}

Crowdsensing approaches maximize their utility when more users are actively participating.
As filing and handling reports inevitably introduces an overhead, achieving a good usability becomes crucial to not deter participation.
Hence, filing a report must be simple and intuitive, even for citizens who are not tech-savvy.
Furthermore, citizens must be presented an easy-to-navigate list of relevant existing reports to facilitate the deduplication of similar reports (\cf~Section~\ref{sec:problem-statement:benefits}).
Finally, the different authorities must be able to receive, review, and respond in a timely manner when reports concerning their respective scope of duties are filed.

Next, we outline how \name currently addresses these challenges and where our early-stage approach requires further improvements.

\section{Accountable Reporting Framework}%
\label{sec:design}

We now outline \name's general approach for realizing moderation capabilities while retaining accountability and its report lifecycle.

\begin{figure}[t]
    \centering
    \includegraphics{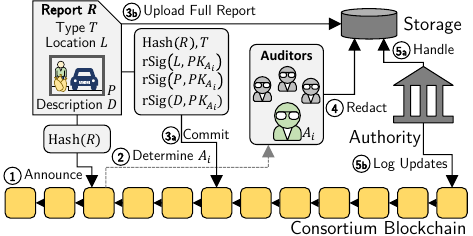}
    \vspace{-0.5em}
    \caption{
        \name's blockchain keeps track of report states.
        Auditors can still moderate reports using redactable signatures.
    }%
    \label{fig:design}
    \vspace{-1em}
\end{figure}

\vspace{-0.2em}
\subsection{General Approach}%
\label{sec:design:idea}

Citizens can file reports using their smartphone.
In our initial design, each report is a tuple \(R = (T, L, P, D)\), where \(T\) is the issue's \emph{type}, \(L\) is a location, \(P\) is a picture of the issue, and \(D\) is a free-text description.
The type is chosen from a fixed set of categories, such that \((T, L)\) describes which authority is responsible for handling the report.

As street issues are a public concern, \name must establish accountability (\cf~Section~\ref{sec:problem-statement:benefits}) but also provide moderation capabilities at the same time to be able to react to wrong or illicit information (\cf~Section~\ref{sec:problem-statement:challenges}).
To establish accountability, \name relies on a consortium blockchain where citizens and authorities log all actions related to announcing or handling issues.
\name further realizes the required moderation by installing a set of semi-trusted \emph{auditors} who vet and potentially redact incoming reports before they are published.
Here, \name relies on \emph{redactable signatures}~\cite{2002_johnson_redact}, which allow dedicated parties to alter a signature's original message without invalidating the signature.
This way, the auditors can redact reports without affecting the immutability of on-chain data.

\vspace{-0.2em}
\subsection{Report Lifecycle}%
\label{sec:design:lifecycle}

In the following, we provide further details about the lifecycle of any report, which consists of five steps as shown by Figure~\ref{fig:design}:

First, the citizen \stepnumberspace{1}{announces} the report by submitting \(\text{Hash}(R)\) to the blockchain.
The announcement does not disclose information about \(R\) but enables the citizen to prove its acceptance later on.
\name can use this input and other blockchain information to pseudorandomly \stepnumberspace{2}{determine} a responsible auditor \(A_i\)~\cite{2020_matzutt_anonboot}.
Afterward, the citizen can select the public key for \(A_i\) and \stepnumberspace[a]{3}{commit} to \(R\) by submitting redactable signatures for \(L\), \(P\), and \(D\) to the blockchain.
Simultaneously, the citizen \stepnumberspace[b]{3}{uploads} the full report to a dedicated storage.
Next, the auditor can optionally redact \(R\), yielding \(R^\prime\), without invalidating the on-chain signatures.
If the auditor performs unwarranted redactions, the citizen can release the original report as evidence to resolve this dispute.
Finally, the report is released to the public and the responsible authority, as determined by \((T, L)\), is asked to \stepnumberspace[a]{5}{handle} the report while \stepnumberspace[b]{5}{logging} all updates.
Whenever reports must be completely discarded, either because they are not actionable by the authority or they cannot be published, the authority is still required to log a deletion action on the blockchain.

\section{First Feasibility Insights of \name}%
\label{sec:eval}

In this section, we assess the potential feasibility of \name by concluding that it achieves accountability and moderation at the same time and by discussing its overhead and usability implications.

\paragraph{Accountability vs. Moderation}

A citizen initially only submits a hash value of their report \(R\) to the blockchain, \ie its content is not leaked to the responsible authority, which could otherwise have an interest to reject \(R\).
Assuming that the blockchain network is not compromised by a large-scale attacker, its nodes will thus obliviously accept the announcement of \(R\).
From this point onward, the citizen can always publish \(R\) to prove their submission of \(R\) and its initial content to third parties.
Hence, the responsible authority has to react to the report.
Using redactable signatures further only allows the auditor to remove, but not alter, parts of the report~\cite{2002_johnson_redact}.
In the future, \name must further be hardened against other forms of unwanted user behavior, such as spamming reports.

\begin{figure}[t]
    \centering
    \vspace{.35em}
    \begin{subfigure}{0.32\linewidth}
        \centering
        \includegraphics{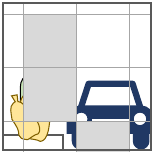}
        \vspace{-.35em}
        \caption{Grid-based coarse }
        \label{fig:redaction-coarse}
    \end{subfigure}
    \hfill
    \begin{subfigure}{0.32\linewidth}
        \centering
        \includegraphics{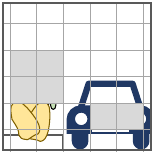}
        \vspace{-.35em}
        \caption{Grid-based fine}
        \label{fig:redaction-fine}
    \end{subfigure}
    \hfill
    \begin{subfigure}{0.32\linewidth}
        \centering
        \includegraphics{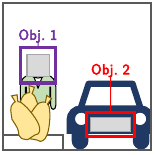}
        \vspace{-.35em}
        \caption{Object-based}
        \label{fig:redaction-object}
    \end{subfigure}
    \vspace{-0.7em}
    \caption{
        Chunking affects the performance and accuracy.
    }
    \label{fig:redaction}
    \vspace{-1em}
\end{figure}

\paragraph{Overhead}

Keeping track of reports and potential redactions necessarily inflicts storage and processing overhead.
However, we argue that this overhead is reasonable given the achievable benefits.
Namely, storing only references to reports on-chain helps reducing the blockchain's growth rate and the redactable signatures per report comprise the bulk of the on-chain data.
The redactable-signature scheme by Johnson \etal~\cite{2002_johnson_redact} assumes that the input message consists of multiple chunks and then yields signatures that grow logarithmically depending on message and chunk lengths and linearly depending on the number of redacted chunks.
Furthermore, as illustrated by Figures~\ref{fig:redaction-coarse} and~\ref{fig:redaction-fine}, the chunk size also determines the granularity available to the auditor, which could ultimately lead to necessary redactions that shadow a reported issue.
Ideally, auditors were able to identify different objects of the report (Figure~\ref{fig:redaction-object}) without large performance penalties.
The assessment of available redaction schemes~\cite{2017_bilzhause_redactable_position} as well as the chunking mechanisms are thus important future work to improve \name.

\paragraph{Usability}

Having one decentralized backend for reporting street problems proves beneficial for mobile citizens as they can access \name from every supporting municipality.
Establishing consensus about reported issues despite this decentralization further helps the desirable deduplication of reports because citizens can be presented a more reliable list of potentially similar issues, \ie of the same type in the same proximity.
Extending this review process with social features such as up-voting or commenting registered reports then presents a valuable input for the prioritization on the authorities' end.
However, \name's utility depends on how intuitive and easy filing and reviewing reports is and maximizing this utility in the future is crucial for increasing \name's acceptance.

In conclusion, \name's approach of combining blockchain-backed accountability with moderation capabilities due to redactable signatures is promising to facilitate the decentralized, accountable, and usable reporting of street issues by mobile citizens.

\section{Conclusion}%
\label{sec:conclusion}

We proposed the accountable city report platform (\name) to increase the transparency of municipalities handling street problems such as potholes or illegal trash dumps.
We identified a combination of blockchain-recorded meta information and redactable digital signatures as promising to establish this transparency, and thereby accountability, while retaining important moderation capabilities required for operating \name in the public.
Our initial assessment suggests that we can extend \name to become a valuable tool for the modern city management.
We are looking forward to exploring the requirements for realizing \name, and potential further use cases of its concept, in greater detail in the future.

\printbibliography

\end{document}